\documentstyle[12pt,epsfig]{article}

\date{}

\begin{document}

\title{{\LARGE\sf Critical Behavior of the Kramers Escape Rate in Asymmetric Classical
Field Theories}}
\author{
{\bf D. L. Stein}\thanks{Partially supported by the 
National Science Foundation under grant PHY-0099484.}\\
{\small \tt dls\,@\,physics.arizona.edu}\\
{\small \sl Departments\ of Physics and Mathematics}\\
{\small \sl University of Arizona}\\
{\small \sl Tucson, AZ 85721, USA}
}

\maketitle

\begin{abstract}

We introduce an asymmetric classical Ginzburg-Landau model in a bounded
interval, and study its dynamical behavior when perturbed by weak
spatiotemporal noise.  The Kramers escape rate from a locally stable state
is computed as a function of the interval length.  An asymptotically sharp
second-order phase transition in activation behavior, with corresponding
critical behavior of the rate prefactor, occurs at a critical length
$\ell_c$, similar to what is observed in symmetric models.  The weak-noise
exit time asymptotics, to both leading and subdominant orders, are analyzed
at all interval lengthscales.  The divergence of the prefactor as the
critical length is approached is discussed in terms of a crossover from
non-Arrhenius to Arrhenius behavior as noise intensity decreases.  More
general models without symmetry are observed to display similar behavior,
suggesting that the presence of a ``phase transition'' in escape behavior
is a robust and widespread phenomenon.

\end{abstract}

{\bf KEY WORDS:} Fokker-Planck equation, non-Arrhenius behavior, stochastic
escape problem, stochastic exit problem, stochastically perturbed dynamical
systems, spatiotemporal noise, droplet nucleation, fluctuation determinant,
instanton, false vacuum, stochastic Ginzburg-Landau models.

\small
\renewcommand{\baselinestretch}{1.25}
\normalsize


\section{Introduction}
\label{sec:intro}

Noise-induced transitions between locally stable states of spatially
extended systems are responsible for a wide range of physical
phenomena$^{\footnotesize{\cite{GarciaOjalvo99}}}$.  In classical systems,
where the noise is typically (but not necessarily) of thermal origin, such
phenomena include homogeneous nucleation of one phase inside
another$^{\footnotesize{\cite{Langer67}}}$, micromagnetic domain
reversal$^{\footnotesize{\cite{Braun93,Boerner98,Brown00}}}$, pattern
nucleation in electroconvection$^{\footnotesize{\cite{Bisang98andTu97}}}$
and other non-equilibrium systems$^{\footnotesize{\cite{Cross93}}}$,
transitions in hydrogen-bonded
ferroelectrics$^{\footnotesize{\cite{Dikande97}}}$, dislocation motion
across Peierls barriers$^{\footnotesize{\cite{Gorokhov97}}}$,
instabilities of metallic nanowires$^{\footnotesize{\cite{Buerkiinprep}}}$,
and others.  In quantum systems, the problem of tunneling between
metastable states is formally similar, and problems of interest include
decay of the false vacuum$^{\footnotesize{\cite{Coleman77andCallan79}}}$
and metastable states in general$^{\footnotesize{\cite{Affleck81}}}$,
anomalous particle production$^{\footnotesize{\cite{Frost99}}}$, and
others.

The modern approach to these problems, beginning with the work of Langer on
classical systems$^{\footnotesize{\cite{Langer67}}}$ and Coleman and Callan
on quantum systems$^{\footnotesize{\cite{Coleman77andCallan79}}}$,
considered systems of infinite spatial extent (for a review, see
Schulman$^{\footnotesize{\cite{Schulman81}}}$).  In certain systems,
however, finite size may lead to important modifications, and in some
instances qualitatively new behavior.  Approaches to noise-induced
transitions between stable states in finite systems modelled by nonlinear
field equations have been investigated by a number of
authors$^{\footnotesize{\cite{Faris82,Martinelli89,Chudnovsky92,McKane95,Kuznetsov97}}}$.

In a recent paper$^{\footnotesize{\cite{MS01}}}$, Maier and Stein studied
the effects of weak white noise on a bistable classical system of finite
size whose zero-noise dynamics are governed by a symmetric Ginzburg-Landau
$\phi^4$ double-well potential.  Their surprising result was the uncovering
of a type of {\it second-order phase transition\/} in activation behavior
at a critical value $L_c$ of the system size.  That a crossover in
activation behavior must take place is clear from both simple physical and
mathematical arguments (cf.~Sec.~\ref{sec:model}).  What is not so obvious
is that the crossover is an asymptotically sharp, second-order phase
transition in the limit of low noise. The change of behavior arises from a
bifurcation of the transition state, from a zero-dimensional (i.e.,
constant) configuration below $L_c$, to a spatially varying (degenerate)
pair of ``periodic instantons'' above $L_c$.

The quantitative effects of the transition are significant.  In the
weak-noise limit, the activation rate is given by the Kramers formula
$\Gamma\sim \Gamma_0\exp(-\Delta W/\epsilon)$, where $\epsilon$~is the
noise strength, $\Delta W$ the activation barrier, and $\Gamma_0$ the rate
prefactor.  The barrier $\Delta W$ is interpreted as the height, in
dimensionless energy units, of the transition state, and by analogy with
chemical kinetics, the exponential falloff of the rate is often called
`Arrhenius behavior'.  The dependence on system size $L$ of $\Delta W$
changes qualitatively at $L_c$.  Also, the rate prefactor $\Gamma_0$ {\it
diverges\/} as $L_c$ is approached both from above and below.  Precisely at
$L_c$, $\Gamma_0$ becomes $\epsilon$-dependent in such a way that it
diverges as $\epsilon\to 0$.  This is `non-Arrhenius' behavior.  (For
boundary conditions that give rise to a zero mode, such as periodic, there
is in addition a noise dependence of $\Gamma_0$ above $L_c$, and the
prefactor divergence as $L\to L_c^+$ may be affected.)

Given the increasingly anomalous behavior of the escape rate as $L_c$ is
approached from either side, a few words should be said about the domain of
validity of the Kramers formula $\Gamma\sim \Gamma_0\exp(-\Delta
W/\epsilon)$ for the escape rate, which displays Arrhenius behavior with
both $\Gamma_0$ and $\Delta W$ independent of $\epsilon$.  Strictly
speaking, this formula is {\it asymptotically\/} valid, that is, only in
the limit $\epsilon\to 0$.  In a looser sense, the formula can often be
applied to physical situations when the noise strength $\epsilon$ is both
small compared to $\Delta W$, and so that the prefactor $\Gamma_0$ is small
compared to $\exp(-\Delta W/\epsilon)$.  These represent minimal
requirements; in all cases applications need to be made with care.  For a
fuller discussion on these and related issues, see \cite{HTB90}.  For the
models presented here, a discussion of the regions of validity of all
derived formulae will be presented in Sec.~\ref{subsec:divergence}.

A question that naturally arises is whether this critical behavior is
generic: could it depend on special features of the potential studied in
\cite{MS01}, in particular its $\phi\to -\phi$ symmetry?  It was
noted$^{\footnotesize{\cite{MS01}}}$ that in more complicated
models$^{\footnotesize{\cite{Kuznetsov97}}}$, the transition may become
first-order; in others, it could conceivably disappear altogether.  The
purpose of this paper, however, is to provide support for the claim that
the transition found in \cite{MS01} is at least not confined to models with
$\phi\to -\phi$ symmetry; that it should in fact appear in a wide range of
models and corresponding physical situations.  To support this, a
nonsymmetric $\phi^3$ model will be studied and solved, and a second-order
transition similar to that found in \cite{MS01} will be uncovered.  This
will be followed by a brief discussion of general nonsymmetric
Ginzburg-Landau models with smooth polynomial potentials up to degree four,
and it will be argued that this second-order transition should appear in
typical representations of these models.

\section{The Model}
\label{sec:model}

We consider on $[-L/2,L/2]$ a classical field $\phi(x,t)$ subject to the potential
\begin{equation}
\label{eq:field}
V(\phi) = -\alpha\phi + {1\over 3}\gamma\phi^3
\end{equation}
as shown in Fig.~1.

\begin{figure}
\vskip 0.5in
\centerline{\epsfig{file=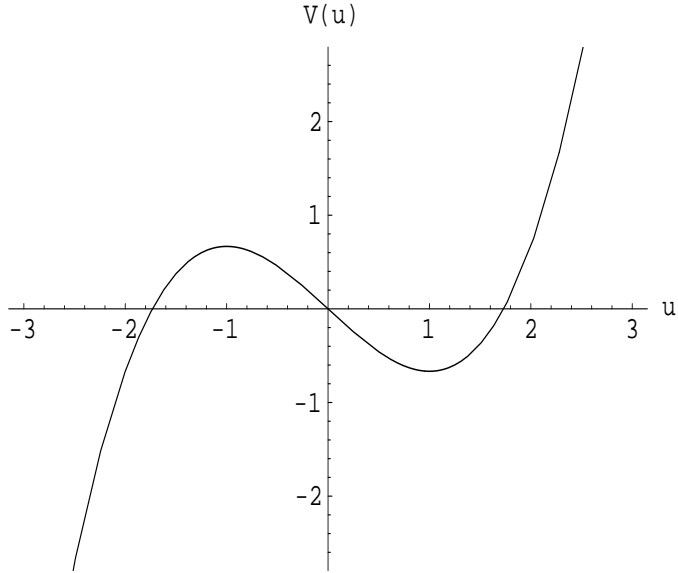,width=3.5in,height=3.0in}}
\renewcommand{\baselinestretch}{1.0} 
\small
\caption{Potential for the reduced field $u(x,t)$.}
\label{fig:potential}
\end{figure}
\renewcommand{\baselinestretch}{1.25} 
\normalsize 

The time evolution of the field is governed by the stochastic, overdamped
Ginzburg-Landau equation
\begin{equation}
\label{eq:Langevin}
\dot\phi = \kappa\phi'' + \alpha - \gamma\phi^2 + \epsilon^{1/2}\xi(x,t),
\end{equation}
where $\xi(x,t)$ is white noise, satisfying
$\langle\xi(x_1,t_1)\xi(x_2,t_2)\rangle=\delta(x_1-x_2)\delta(t_1-t_2)$.
The zero-noise dynamics satisfy $\dot\phi=-\delta{\cal H}/\delta\phi$, with the energy
functional
\begin{equation}
\label{eq:energy}
{\cal H}[\phi]\equiv
\int_{-L/2}^{L/2} \left[{1\over 2}\kappa(\phi')^2 -\alpha\phi + {1\over 3}\gamma\phi^3\right]
\,dz.
\end{equation}

Scaling out the various constants by introducing the variables
$x=\Big[(\alpha\gamma)^{1/4}/\kappa^{1/2}\Big]z$,
$u=\sqrt{\gamma/\alpha}\phi$,
and $E_0=\kappa^{1/2}\alpha^{5/4}/\gamma^{3/4}$ yields
\begin{equation}
\label{eq:scaledenergy}
{\cal H}[u]/E_0=\int_{-\ell/2}^{\ell/2} \left[{1\over 2}(u')^2  -u + {1\over
3}u^3\right]
\,dx
\end{equation}
where $\ell=\Big[(\alpha\gamma)^{1/4}/\kappa^{1/2}\Big]L$.

It is already clear that a crossover in activation behavior must occur.  In
the limit $\ell\to 0$ the gradient term in the integrand of the energy in
Eq.~(\ref{eq:scaledenergy}) will diverge for a nonuniform state; while for
$\ell\to\infty$ the $V(u)$ term will diverge for a uniform state.  In this
paper we will employ periodic boundary conditions throughout, and it is
clear that there must be a crossover from a uniform to a nonuniform
transition state as $\ell$ increases from 0.  Physically, the crossover
arises from a competition between the bending and bulk energies of the
transition state.

This crossover will be analyzed in succeeding sections; we will see that it
corresponds to an asymptotically sharp, second-order phase transition in
the activation rate.  Both stable and transition states are
time-independent solutions of the zero-noise Ginzburg-Landau equation, that
is, they are extremal states of ${\cal H}[\phi]$, satisfying
\begin{equation}
\label{eq:extrema}
u''=-1+u^2\, .
\end{equation}

As already noted, we will assume periodic boundary conditions throughout.
So there is a uniform stable state $u_s=+1$, and a uniform unstable state
$u_u=-1$.  In the next section we will see that the latter is the
transition state for $\ell<\ell_c=\sqrt{2}\pi$.  At $\ell_c$ a transition
occurs, and above it the transition state is nonuniform.

\section{The Transition State}
\label{sec:instanton}

\begin{figure}
\vskip 0.5in
\centerline{\epsfig{file=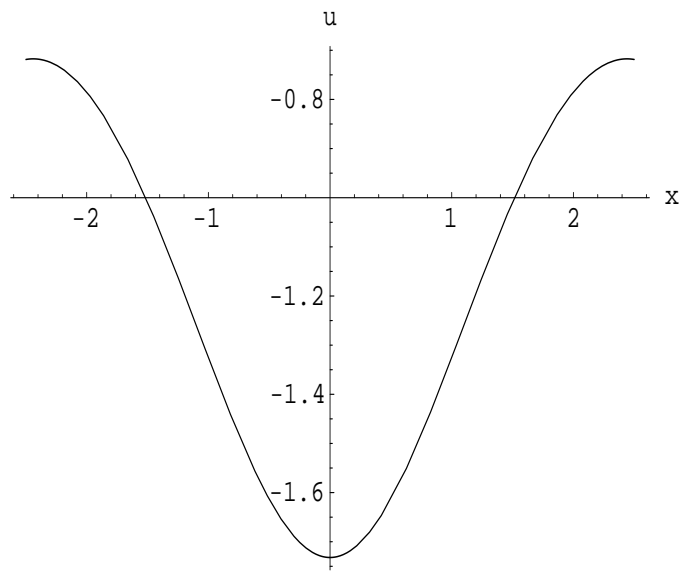,width=3.5in,height=3.0in}}
\renewcommand{\baselinestretch}{1.0} 
\small
\caption{The instanton state $u_{{\rm inst},m}(x)$ for $m=1/2$.}
\label{fig:dn}
\end{figure}

Following the notation of \cite{MS01}, we denote by
$u_{{\rm inst},m}(x)$ the spatially varying, time-independent solution
(``instanton state'') to the zero-noise extremum condition
Eq.~(\ref{eq:extrema}), for any $m$ in the range $0\le m\le 1$.  The
instanton state is (see Fig.~\ref{fig:dn})
\begin{equation}
\label{eq:instanton}
u_{{\rm inst},m}(x,x_0) = {(2-m)\over\sqrt{m^2-m+1}} -
{3\over\sqrt{m^2-m+1}}{\rm dn}^2\left[{x-x_0\over \sqrt{2}(m^2-m+1)^{1/4}} \Big| m\right]\, ,
\end{equation}
where dn$(\cdot|m)$ is the Jacobi elliptic function with parameter $m$,
whose half-period equals ${\bf K}(m)$, the complete elliptic integral of the
first kind$^{\footnotesize{\cite{AS}}}$.  Accordingly, imposition of the
periodic boundary condition yields a relation between $\ell$ and $m$:
\begin{equation}
\label{eq:pbc}
\ell=2\sqrt{2}(m^2-m+1)^{1/4}{\bf K}(m)\, .
\end{equation}

\renewcommand{\baselinestretch}{1.25} 
\normalsize 

\begin{figure}
\vskip 0.5in
\centerline{\epsfig{file=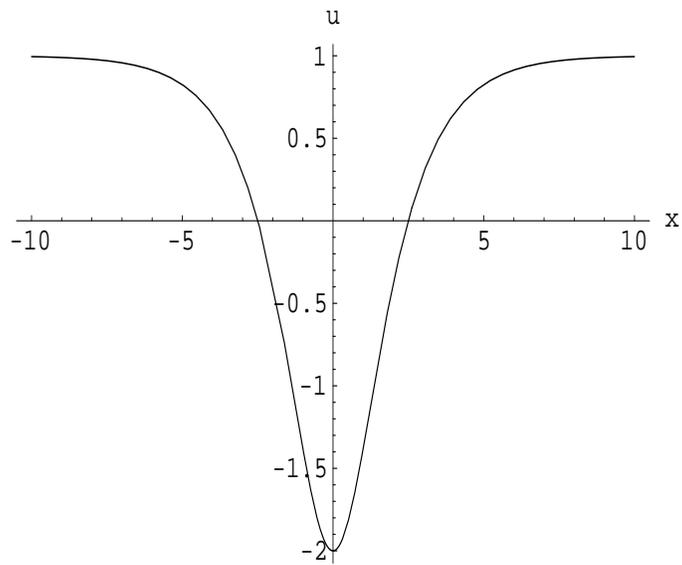,width=3.5in,height=3.0in}}
\renewcommand{\baselinestretch}{1.0} 
\small
\caption{The instanton state $u_{{\rm inst},1}(x)$ on the infinite line.}
\label{fig:sech}
\end{figure}

The minimum length that can accommodate this condition is
$\ell_c=\sqrt{2}\pi$, corresponding to $m=0$.  In this limit,
${\rm dn}(x|0)=1$, and the instanton state reduces to the uniform unstable state
$u_u=-1$.  As $m\to 1^-$, $\ell\to\infty$, and the instanton state becomes
\begin{equation}
\label{eq:sech}
u_{{\rm inst},1}(x)=1-3{\rm sech}^2\left({x-x_0\over\sqrt{2}}\right)\,
\end{equation}
as shown in Fig.~\ref{fig:sech}.  As we will see, the instanton given by
Eq.~(\ref{eq:instanton}) is a saddle, or transition state, above $\ell_c$;
it will be seen to have one unstable direction (in addition to a zero mode
resulting from translational symmetry).  Physically, it can be thought of
as a pair of domain walls, each of which separates the two uniform states
over a region of finite extent.
\renewcommand{\baselinestretch}{1.25} 
\normalsize 

\begin{figure}
\centerline{\epsfig{file=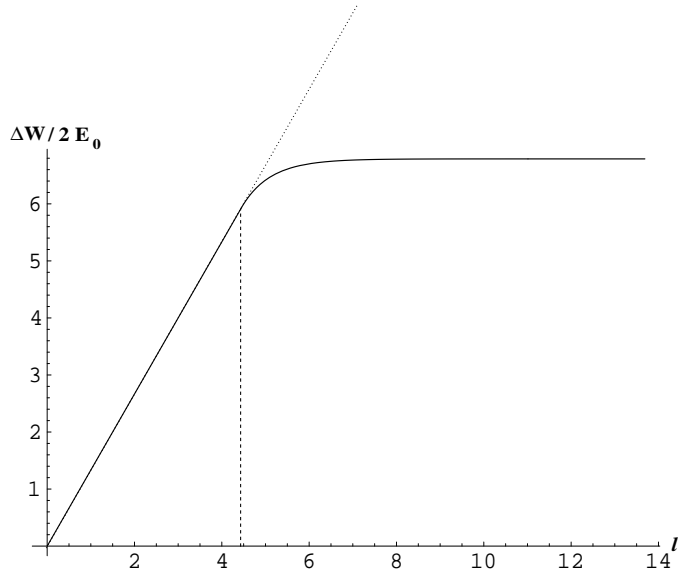,width=3.5in,height=3.0in}}
\renewcommand{\baselinestretch}{1.0} 
\small
\caption{The activation barrier $\Delta W/2E_0$ for periodic boundary
conditions (solid line).  The dashed line indicates the crossover from the
uniform transition state to the instanton transition state at $\ell_c$, and
the dotted line an extension to the region $\ell>\ell_c$ of the activation
energy corresponding to the uniform state.}
\label{fig:activation}
\end{figure}

For small but nonzero noise strength the leading-order asymptotics in the
escape rate from the metastable well are governed by the energy difference
between the transition state, which is $u_u=-1$ below $\ell_c$ and $u_{{\rm
inst},m}(x)$ above, and the stable state $u_s=+1$.  (A stability analysis
justifying these identifications will be given in
Secs.~\ref{subsec:subcrit} and \ref{subsec:supercrit}.)  In the Kramers
formula in Sec.~\ref{sec:intro}, the activation barrier $\Delta W$, which
governs exponential dependence of the escape rate on the noise strength,
equals twice this energy difference.  So below $\ell_c$, $\Delta
W/2E_0=(4/3)\ell$.  Above $\ell_c$, we find
\begin{eqnarray}
\label{eq:barrier}
{\Delta W\over 2E_0} = \left({\cal H}[u_{{\rm inst},m}(x)]-{\cal H}[u_s=1]\right)=
\left[{2-3m-3m^2+2m^3\over 3(m^2-m+1)^{3/2}}+{2\over 3}\right]\ell\\
+{12\sqrt{2}\over 5(m^2-m+1)^{1/4}}
\left[2{\bf E}(m)-{(2-m)(1-m)\over(m^2-m+1)}{\bf K}(m)\right]\, ,
\end{eqnarray}
where ${\bf E}(m)$ is the complete elliptic integral of the second
kind$^{\footnotesize{\cite{AS}}}$.  The activation barrier for the entire
range of $\ell$ is shown in Fig.~\ref{fig:activation}.  As $\ell\to\infty$,
$\Delta W/2E_0\to 24\sqrt{2}/5$; this value is simply the energy of a pair
of domain walls.

\renewcommand{\baselinestretch}{1.25} 
\normalsize 

\section{Rate prefactor}
\label{sec:prefactor}

For an overdamped multidimensional system driven by white noise, the rate
prefactor $\Gamma_0$ can be computed as
follows$^{\footnotesize{\cite{HTB90,MM}}}$ (see also
\cite{Langer67,Schulman81}).  As in \cite{MS01}, let ${\bf\varphi}_s$
denote the stable state, and let ${\bf\varphi}_u$ denote the transition
state; it will be assumed (as is the case here) that this state has a
single unstable direction.  Consider a small perturbation ${\bf\eta}$ about
the stable state, i.e., ${\bf\varphi}={\bf\varphi}_s+{\bf\eta}$.  Then to
leading order $\dot{\bf\eta}=-{\bf\Lambda}_s{\bf\eta}$, where
${\bf\Lambda}_s$ is the linearized zero-noise dynamics at ${\bf\varphi}_s$.
Similarly ${\bf\Lambda}_u$ is the linearized zero-noise dynamics around
${\bf\varphi}_u$. Then$^{\footnotesize{\cite{HTB90,MM}}}$
\begin{equation}
\label{eq:prefactor}
\Gamma_0 = 
\frac1{2\pi}
\sqrt{\left|\frac{\det{\bf\Lambda}_s}{\det{\bf\Lambda}_u}\right|}
\,\,\left|\lambda_{u,1}\right|,
\end{equation}
where $\lambda_{u,1}$ is the only negative eigenvalue of~${\bf\Lambda}_u$,
corresponding to the direction along which the optimal escape
trajectory approaches the transition state. In general, the determinants in
the numerator and denominator of Eq.~(\ref{eq:prefactor}) separately
diverge: they are products of an infinite number of eigenvalues with
magnitude greater than one.  However, their {\it ratio\/}, which can be
interpreted as the limit of a product of individual eigenvalue quotients,
is finite.

\subsection{$\ell<\ell_c$}
\label{subsec:subcrit}

In this regime, both the stable and transition states are uniform,
allowing for a straightforward determination of $\Gamma_0$ by direct
computation of the determinants.  Using reduced variables, the stable state
$u_s=+1$ and the transition state $u_u=-1$.  Linearizing around the stable
state gives
\begin{equation}
\label{eq:us}
\dot\eta=-\hat\Lambda[u_s]\eta=-(-d^2/dx^2+2)\eta\, ,
\end{equation}
and similarly
\begin{equation}
\label{eq:uu}
\dot\eta=-\hat\Lambda[u_u]\eta=-(-d^2/dx^2-2)\eta\, 
\end{equation}
about the transition state.  The spectrum of eigenvalues corresponding
to $\hat\Lambda[u_s]$ is
\begin{equation}
\label{eq:stablespectrum}
\lambda_n^s=2+{4\pi^2n^2\over\ell^2}\qquad\qquad\qquad
n=0,\pm 1,\pm 2\ldots
\end{equation}
and the eigenvalues corresponding to $\hat\Lambda[u_u]$ are
\begin{equation}
\label{eq:unstablespectrum}
\lambda_n^u=-2+{4\pi^2n^2\over\ell^2}\qquad\qquad\qquad n=0,\pm 1,\pm 2\ldots\, .
\end{equation}

This simple linear stability analysis justifies the claims that $u_s$ is a
stable state and $u_u$ a transition state, or saddle point.  Over the
interval $[0,\ell_c)$ all eigenvalues of $\hat\Lambda[u_s]$ are positive,
while all but one of $\hat\Lambda[u_u]$ are.  Its single negative
eigenvalue $\lambda_0^u=-2$ is independent of $\ell$, and the corresponding
eigenfunction, which is spatially uniform, is the direction in
configuration space along which the optimal escape path approaches $u_u$.

Putting everything together, we find
\begin{eqnarray}
\label{eq:g0-}
\Gamma_0&=&{1\over\pi}\sqrt{\left|{\prod_{n=-\infty}^\infty(2+{4\pi^2n^2\over\ell^2})\over
\prod_{n=-\infty}^\infty(-2+{4\pi^2n^2\over\ell^2})}\right|}\nonumber \\
&=&{1\over\pi}{\sinh(\ell/\sqrt{2})\over\sin(\ell/\sqrt{2})}\, ,
\end{eqnarray}
which diverges at $l_c=\sqrt{2}\pi$ as expected; in this limit,
$\Gamma_0\sim {\rm const}\times (\ell_c-l)^{-1}$.  The divergence arises
from the vanishing of the pair of eigenvalues $\lambda^u_{\pm 1}$ as
$\ell\to\ell_c^-$ (each eigenvalue contributing a factor
$(\ell_c-\ell)^{-1/2}$).  This indicates the appearance of a pair of soft
modes, resulting in a transversal instability of the optimal escape
trajectory as the saddle point is approached.

\subsection{$\ell>\ell_c$}
\label{subsec:supercrit}

Computation of the determinant quotient in Eq.~(\ref{eq:prefactor}) is less
straightforward when the transition state is nonconstant.  This occurs when
$\ell>\ell_c$, where the transition state $u_u$ is given by
Eq.~(\ref{eq:instanton}), and its associated linearized evolution operator
is
\begin{equation}
\label{eq:uuop}
\hat\Lambda[u_u]=-{d^2\over
dx^2}+{2(2-m)\over\sqrt{m^2-m+1}}-{6\over\sqrt{m^2-m+1}}\ {\rm
dn}^2\left[{x-x_0\over\sqrt{2}(m^2-m+1)^{1/4}}\Big|m\right]\, .
\end{equation}
Evaluation of $\Gamma_0$ therefore requires determination of the eigenvalue
spectrum of $\hat\Lambda[u_u]$ with periodic boundary conditions.

An additional complication follows from the infinite translational
degeneracy of the instanton state (i.e., invariance with respect to choice
of $x_0$).  This implies a soft collective mode in the linearized dynamical
operator $\hat\Lambda[u_u]$ of Eq.~(\ref{eq:uuop}), resulting in a zero
eigenvalue.  Removal of this zero eigenvalue can be achieved with the
McKane-Tarlie regularization procedure$^{\footnotesize{\cite{McKane95}}}$
for functional determinants.

That procedure is implemented as follows (see \cite{McKane95} for details).
Let $y_1(x,x_0;m)$ and $y_2(x,x_0;m)$ denote two linearly independent
solutions of $\hat\Lambda[u_u]y_i=0$, $i=1,2$.  Let $\det'\hat\Lambda$ refer to the
functional determinant of the operator $\hat\Lambda$ with the zero
eigenvalue removed.  Then, with periodic boundary conditions, it is
formally the case that
\begin{equation}
\label{eq:regdet}
{\det'\hat\Lambda\over\langle y_1|y_1\rangle}
={y_2(z+\ell)-y_2(z)\over y_1(z)\det{\bf H}(z)}
\end{equation}
where $z$ is arbitrary,
$\langle y_1|y_1\rangle=\int_{-\ell/2}^{\ell/2}dx\ y_1^2(x,0;m)$ is the
square of the norm of the zero mode and $\det{\bf
H}(z)=\dot y_2(z)y_1(z)-\dot y_1(z)y_2(z)$ is the Wronskian.
(The expression (\ref{eq:regdet}) is meaningful only as part of a
determinant quotient; see below.)

Solutions $y_1$ and $y_2$ of $\hat\Lambda[u_u]y_i=0$ can be found by
differentiating the instanton solution (\ref{eq:instanton}) with respect to
$x_0$ and $m$, respectively; i.e., $y_1(x,x_0;m)=\partial u_{{\rm
inst},m}(x,x_0;m)/\partial x_0$ and $y_2(x,x_0;m)=\partial u_{{\rm
inst},m}(x,x_0;m)/\partial m$.  The homogeneous solutions are then
\begin{equation}
\label{eq:y1} 
y_1(x,x_0;m)=-{6m\beta(m)^3\over\sqrt{2}}{\rm
sn}\left[{\beta(m)(x-x_0)\over\sqrt{2}}\Big|m\right] {\rm
cn}\left[{\beta(m)(x-x_0)\over\sqrt{2}}\Big|m\right]{\rm
dn}\left[{\beta(m)(x-x_0)\over\sqrt{2}}\Big|m\right]
\end{equation}
and
\begin{eqnarray}
\label{eq:y2} 
y_2(x,x_0;m)&=&-{3m\beta(m)^6\over 2}+{3\beta(m)^2\over(1-m)}{\rm
sn}^2\left[{\beta(m)(x-x_0)\over\sqrt{2}}\Big|m\right]{\rm
dn}^2\left[{\beta(m)(x-x_0)\over\sqrt{2}}\Big|m\right]\nonumber\\
&+&{3(2m-1)\beta(m)^6\over 2}{\rm
dn}^2\left[{\beta(m)(x-x_0)\over\sqrt{2}}\Big|m\right]\nonumber\\
&+&3\beta(m)^2{\rm sn}\left[{\beta(m)(x-x_0)\over\sqrt{2}}\Big|m\right]
{\rm cn}\left[{\beta(m)(x-x_0)\over\sqrt{2}}\Big|m\right]{\rm
dn}\left[{\beta(m)(x-x_0)\over\sqrt{2}}\Big|m\right]\nonumber\\
&\times&\left\{{(2-m)\beta(m)^5(x-x_0)\over 2\sqrt{2}}-{{\bf
E}\left({\beta(m)(x-x_0)\over\sqrt{2}}\Big|m\right)\over 1-m}\right\}\, ,
\end{eqnarray}
where $\beta(m)=(m^2-m+1)^{-1/4}$ and ${\bf E}(\cdot|m)$ is the incomplete
elliptic integral of the second kind$^{\footnotesize{\cite{AS}}}$.
Inserting these solutions into Eq.~(\ref{eq:regdet}) yields
\begin{equation}
\label{eq:unstable}
\left|{\det'\hat\Lambda[u_u]\over\langle y_1|y_1\rangle}\right|={(m^2-m+1)^{11/4}\over
9m^2(1-m)}\left[{2{\bf E}(m)\over 1-m}-{(2-m){\bf K}(m)\over m^2-m+1}\right]\, .
\end{equation}

Using a similar procedure (see the Appendix of \cite{McKane95}), we find
the corresponding numerator for the determinant ratio in
Eq.~(\ref{eq:prefactor}) to be
\begin{equation}
\label{eq:stable}
\det\hat\Lambda[u_s]=4\sinh^2(\ell/\sqrt{2})\, ,
\end{equation}
consistent with the numerator of Eq.~(\ref{eq:g0-}), obtained through
direct computation of the eigenvalue spectrum.  We emphasize again, however
(cf.~the discussion above Eq.~(\ref{eq:g0-})), that it is only the {\it
ratio} of the determinants that is sensible, not the individual
determinants themselves: these each diverge for every $\ell$.  In contrast,
the expressions in Eqs.~(\ref{eq:unstable}) and (\ref{eq:stable}) are
well-behaved for all finite $\ell>\ell_c$ ($m>0$), but still separately
diverge in the $\ell\to\infty$ ($m\to 1$) limit.  Nevertheless, here also
the divergences cancel to give
\begin{equation}
\label{eq:inflimit}
\lim_{m\to
1}\sqrt{\left|{\det\hat\Lambda[u_s]\over\det'\hat\Lambda[u_u]}\right|}
=24\sqrt{2}\Big/\sqrt{\langle y_1|y_1\rangle}\, .
\end{equation}

We next compute the eigenvalue $\lambda_{u,1}$ corresponding to the
unstable direction.  With the substitution $w=\beta(m)(x-x_0)/\sqrt{2}$, the
eigenvalue equation $\hat\Lambda[u_u]\eta=\lambda\eta$ becomes
\begin{equation}
\label{eq:Lame'}
d^2\eta/dw^2+12{\rm dn}^2[w|m]\eta={\cal E}\eta
\end{equation}
where ${\cal E}=-2\lambda/\beta(m)^2+4(2-m)$.  Using the
identity$^{\footnotesize{\cite{AS}}}$ ${\rm dn}^2[z|m]=1-m{\rm sn}^2[z|m]$,
we observe that Eq.~(\ref{eq:Lame'}) is the $l=3$ Lam\'e
equation$^{\footnotesize{\cite{WW27,LK00}}}$, a Schr\"odinger equation with
periodic potential of period $2{\bf K}(m)$.  Its Bloch wave spectrum
consists of four energy bands, and its eigenfunctions can be expressed in
terms of {\it Lam\'e polynomials}$^{\footnotesize{\cite{WW27}}}$.  A fuller
discussion, especially for higher $l$-values, is given in \cite{rsminprep};
for our purposes here we do not need to utilize the full machinery of the
Hermite solution (a detailed treatment is given in \cite{WW27}).

It is easy to check that the eigenvector $\eta_{u_1}$ with smallest
eigenvalue is
\begin{equation}
\label{eq:eigenvector}
\eta_{u,1}(x,x_0;m)={\rm
dn}^3\left[{\beta(m)(x-x_0)\over\sqrt{2}}\Big|m\right]
+C(m){\rm dn}\left[{\beta(m)(x-x_0)\over\sqrt{2}}\Big|m\right]\, 
\end{equation}
where $C(m)=-[2(2-m)/5]+(1/5)\sqrt{4m^2-m+1}$, and
\begin{equation}
\label{eq:eigenvalue}
\lambda_{u,1}=-\left[{1\over 2}(2-m)+\sqrt{4m^2-m+1}\right]\beta(m)^2\, ,
\end{equation}
which approaches $-2$ as $m\to 0$ in agreement with the single negative
eigenvalue of Eq.~(\ref{eq:unstablespectrum}).

As noted in \cite{MS01}, the full translational symmetry of the periodic
instanton state in the periodic boundary condition case corresponds to a
`soft mode', resulting in appearance of a zero eigenvalue $\lambda_{u,2}=0$
of the operator $\hat\Lambda[u_u]$.  (The corresponding eigenfunction
$\eta_{u,2}$ is given by $y_1$ in Eq.~(\ref{eq:y1}).)  Physically, this
corresponds to zero translational energy of the instanton; i.e., it can
appear anywhere in the volume (in contrast with, say, Dirichlet boundary
conditions, where the instanton is `pinned'.)  This should result in an
overall factor of $\ell$, indicating that the physical quantity of interest
above $\ell_c$ is the transition rate per unit length.  The general
procedure for including this correction is described by Schulman
\cite{Schulman81}.  (Our case differs by a factor of 2 from his due to the
lack of symmetry in our model.)  The net result is to multiply the
prefactor by
\begin{equation}
\label{eq:correction}
\zeta=\ell\langle y_1|y_1\rangle^{1/2}/\sqrt{\pi\epsilon}\, .
\end{equation}  
The most important qualitative changes are the $\epsilon^{-1/2}$ factor,
leading to a non-Arrhenius transition rate above $\ell_c$, and the effect
on the behavior as $\ell\to\ell_c^+$; both will be discussed in more detail
below.

The above discussion also makes clear that it is not necessary to
separately evaluate $\langle y_1|y_1\rangle$.  For completeness' sake,
however, we present it as well.  Its evaluation is straightforward and
yields
\begin{equation}
\label{eq:eta1norm}
\langle y_1|y_1\rangle={72\sqrt{2}\beta(m)\over 15}{\bf E}(m)-
{36\sqrt{2}\beta(m)^5(m^2-3m+2)\over 15}{\bf K}(m)
\end{equation}
which equals $24\sqrt{2}/5$ in the $m\to 1$ limit.

Putting everything together, we find the rate {\it per unit length\/}
prefactor for $\ell>\ell_c$ to be
\begin{eqnarray}
\label{eq:prefabove}
\Gamma_0/\ell&=&{\epsilon^{-1/2}\over\pi^{3/2}}{3m(1-m)\over(m^2-m+1)^{11/8}}\left[{1\over
2}(2-m)+\sqrt{4m^2-m+1}\right]\nonumber\\ &\times&
\sqrt{\sinh^2\left(2(m^2-m+1)^{1/4}{\bf K}(m)\right)\over
2(m^2-m+1){\bf E}(m)-(2-m)(1-m){\bf K}(m)}\, .
\end{eqnarray}

\begin{figure}[t]
\centerline{\epsfig{file=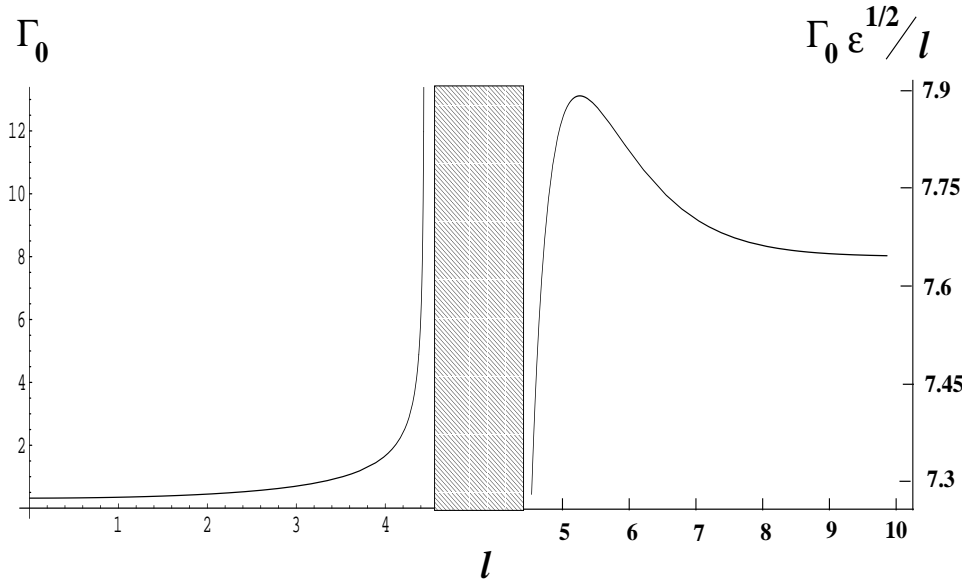,width=5.0in,height=3.0in}}
\renewcommand{\baselinestretch}{1.0} 
\small
\caption{The Kramers rate prefactor $\Gamma_0$.  Because of the qualitative
change in behavior, from Arrhenius to non-Arrhenius, at $\ell_c$, different
vertical scales are used below and above the transition.}
\label{fig:both}
\end{figure}

The prefactor over the entire range of $\ell$ is plotted in
Fig.~\ref{fig:both}.  The prefactor divergence has a critical exponent of 1
as $\ell\to\ell_c^-$.  Above $\ell_c$ the prefactor is non-Arrhenius
everywhere, and the vertical axis is rescaled to account for the singular
$\epsilon^{1/2}$ behavior.  

\renewcommand{\baselinestretch}{1.25} 
\normalsize 

\begin{figure}[t]
\centerline{\epsfig{file=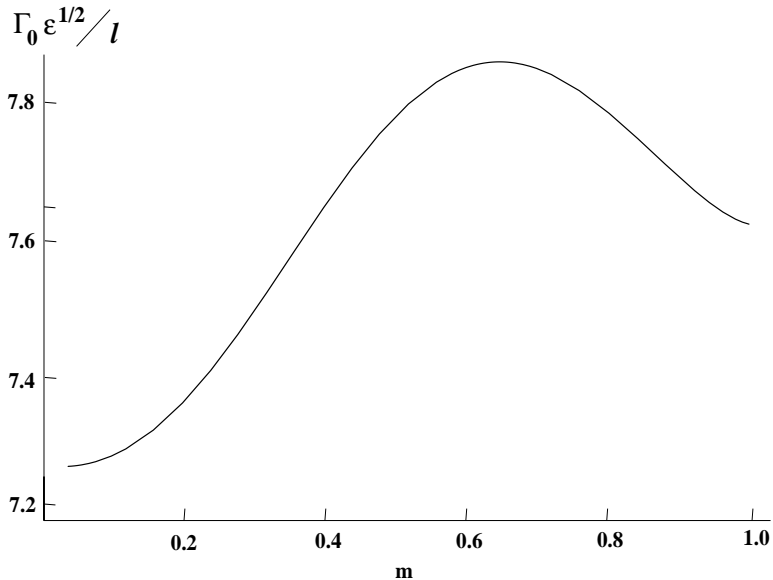,width=4.0in,height=3.0in}}
\renewcommand{\baselinestretch}{1.0} 
\small
\caption{The Kramers rate prefactor $\Gamma_0$ per unit length above
$\ell_c$.}
\label{fig:mplot}
\end{figure}

The rescaled prefactor above $\ell_c$ as a function of $m$ is shown in
Fig.~\ref{fig:mplot}, in order to indicate more clearly the $m\to 0$
($\ell\to\ell_c$) behavior.

\renewcommand{\baselinestretch}{1.25} 
\normalsize 

The behavior of the rate prefactor $\Gamma_0$ for all $\ell>\ell_c$ is
unusual in two ways.  First, it is {\it non-Arrhenius\/} --- that is, it
scales as $\epsilon^{-1/2}$ for all $\epsilon\to 0$.  Second, it does not
formally diverge as $\ell\to\ell_c^+$, as seen in Fig.~\ref{fig:mplot} (in
fact, the divergence is present but `masked', as discussed below).  Both
of these features are consequences of the translation-invariance of the
periodic boundary conditions used here, and would not appear if
translation-noninvariant boundary conditions, such as Dirichlet or Neumann,
are used.  (See, for example, Fig.~3 of \cite{MS01}.)

Such boundary condition-dependent behavior should be distinguished from the
boundary condition-independent formal divergence of the prefactor as the
critical length is approached, as explained below, and of the non-Arrhenius
behavior (to be discussed in Sec.~\ref{subsec:divergence} below) exactly at
the critical length.  By boundary condition-independent, we mean behavior
that is seen in all four of the most commonly used boundary conditions in
this type of problem, namely periodic, antiperiodic, Dirichlet, and
Neumann; all were considered for symmetric quartic potentials in
\cite{MSSPIE}. In the present case of periodic boundary conditions, the
removal of the zero mode that is present for all $\ell>\ell_c$ renormalizes
the prefactor by the factor $\zeta$ in Eq.~(\ref{eq:correction}).  This
renormalization masks the divergence of the determinant ratio as
$\ell\to\ell_c^+$, because the factor $\zeta$ includes the Jacobian of the
transformation$^{\footnotesize{\cite{Schulman81}}}$ from the
translation-invariant normal mode to the variable $x_0$; this in turn
equals the norm $\sqrt{\langle y_1|y_1\rangle}$, which vanishes as $m\to
0$.  The crucial point is that {\it a divergence is still embedded within
the prefactor\/}, in the sense that the square root of the determinant
ratio diverges with a critical exponent of $1/2$ as $\ell\to\ell_c^+$.
Upon closer examination, this arises from the lowest stable eigenvalue,
$\lambda_{u,3}$, of $\det\Lambda'[u_u]$ approaching zero as
$\ell\to\ell_c^+$, in a similar fashion to the eigenvalue behavior below
$\ell_c$ (cf.~$n=\pm 1$ in Eq.~(\ref{eq:unstablespectrum})).  This
eigenvalue and its corresponding eigenfunction $\eta_{u,3}$ are given by
\begin{equation}
\label{eq:eigenvalue3}
\eta_{u,3}(x,x_0;m)={\rm
dn}^3\left[{\beta(m)(x-x_0)\over\sqrt{2}}\Big|m\right]
+C'(m){\rm dn}\left[{\beta(m)(x-x_0)\over\sqrt{2}}\Big|m\right]\, 
\end{equation}
where $C'(m)=-[2(2-m)/5+(1/5)\sqrt{4m^2-m+1}]$, and
\begin{equation}
\label{eq:eigenfunction3}
\lambda_{u,3}=-\left[{1\over 2}(2-m)-\sqrt{4m^2-m+1}\right]\beta(m)^2\, .
\end{equation}

\begin{figure}[t]
\centerline{\epsfig{file=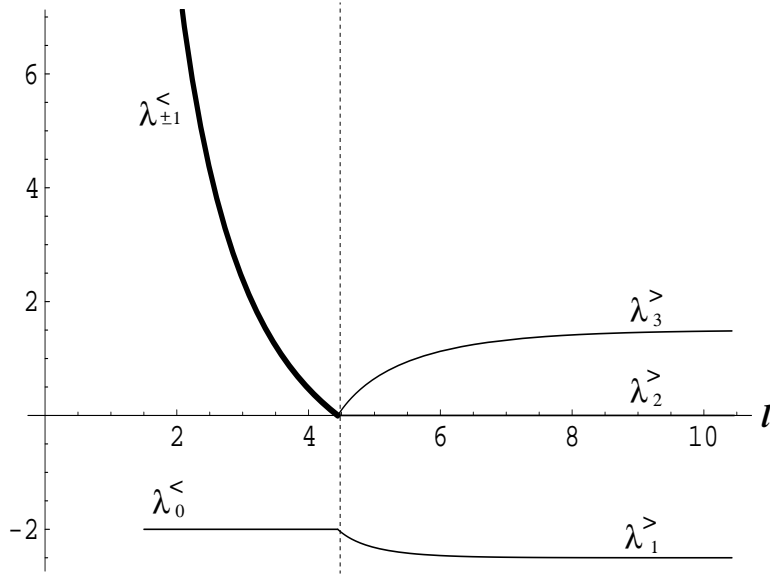,width=4.0in,height=3.0in}}
\renewcommand{\baselinestretch}{1.0} 
\small
\caption{Behavior of the lowest three eigenvalues of the operators
$\hat\Lambda[u_u^<]$ and $\hat\Lambda[u_u^>]$.  The vertical dashed line is
$\ell=\ell_c$.  The eigenvalues $\lambda_n^<$ correspond to $n=0,\pm 1$ of
Eq.~(\ref{eq:unstablespectrum}).  Computation of the three lowest
eigenvalues $\lambda_{1,2,3}^>$ in the $\ell>\ell_c$ region is given in the
text.  The bold curve corresponding to $\lambda_{\pm 1}^<$ reflects its
double degeneracy.}
\label{fig:eigenvalues}
\end{figure}

\begin{figure}[t]
\centerline{\epsfig{file=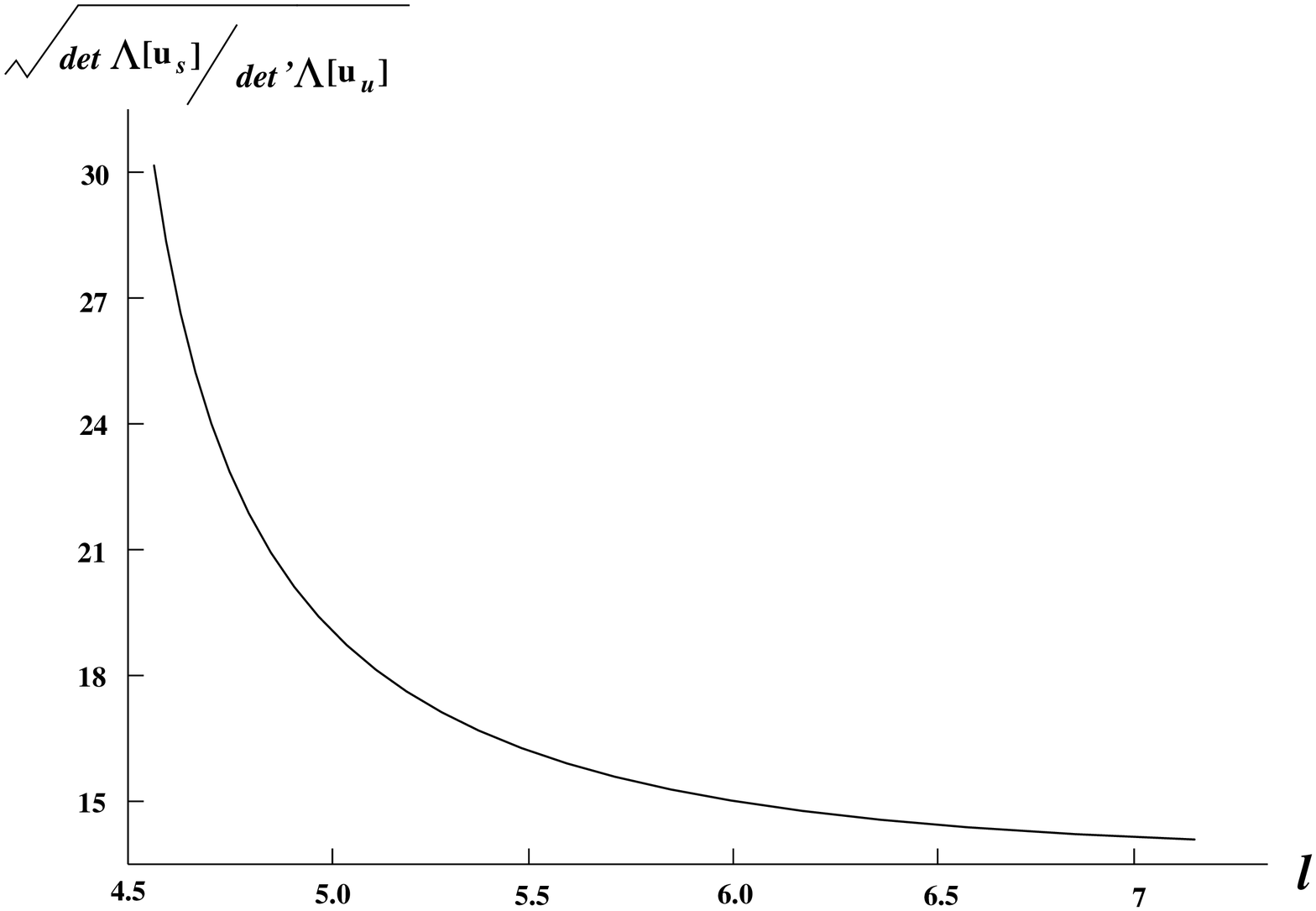,width=4.0in,height=3.0in}}
\renewcommand{\baselinestretch}{1.0} 
\small
\caption{Behavior of the ratio of the eigenvalue spectrum above $\ell_c$.}
\label{fig:ratio}
\end{figure}

Fig.~\ref{fig:eigenvalues} shows the lowest three eigenvalues for the
operators $\hat\Lambda[u_u^<]$ and $\hat\Lambda[u_u^>]$, where $u_u^<$
($u_u^>$) indicates the transition state below (above) $\ell_c$.  This
figure illustrates the evolution of the eigenvalues ($\lambda_{\pm 1}^<$
and $\lambda_3^>$) that control the formal prefactor divergence as $\ell$
passes through $\ell_c$.  We note in particular the merging of the second
and third eigenvalues of $\hat\Lambda[u_u^>]$ as $\ell\to\ell_c^+$,
consistent with the double degeneracy of the corresponding eigenfunction
when $\ell<\ell_c$.  The eigenvalues are everywhere continuous.
Fig.~\ref{fig:ratio} displays the behavior of the full determinant ratio
above $\ell_c$.

\renewcommand{\baselinestretch}{1.25} 
\normalsize 

\subsection{Interpretation of the Prefactor Divergence}
\label{subsec:divergence}

The formal divergence of the Kramers rate prefactor at a critical length
$l_c$ (cf.~Fig.~5) requires interpretation.  It is interesting that a
prefactor divergence was also found$^{\footnotesize{\cite{MS93,MS96}}}$ in
a completely different set of systems, namely spatially {\it homogeneous\/}
(i.e., zero-dimensional) systems {\it out\/} of equilibrium, i.e., in which
detailed balance is not satisfied in the stationary state.  That divergence
arose from an entirely different reason: the appearance of a {\it caustic
singularity} in the vicinity of the most probable exit path as a parameter
in the drift field varied.  The caustic singularity arises from the {\it
unfolding of a boundary catastrophe\/}; a detailed analysis in given in
\cite{MS00}.  In contrast, the problem considered here is that of a
spatially {\it extended\/} system {\it in\/} equilibrium, so no such
singularities can be present.  Moreover, no parameter in the stochastic
differential equations describing the time evolution of the system is being
varied in the case under discussion here; rather, the variation is in the
length of the interval on which the field is defined.  The `phase
transitions' in the stochastic exit problem in the two classes of systems
are therefore physically unrelated.

What does it mean for the prefactor to (formally) diverge?  In fact, at no
lengthscale is the {\it true prefactor\/} infinite, for any $\epsilon>0$.
Indeed, given that the analysis presented here is, strictly speaking, valid
only asymptotically as $\epsilon\to 0$, {\it the escape rate is always
small\/} where the above results are applicable.  What the formal
divergence of the prefactor {\it does\/} mean is that the escape behavior
becomes increasingly anomalous as $\ell_c$ is approached, and it is
asymptotically {\it non-Arrhenius\/} exactly at $\ell_c$.

That is, when $\ell=\ell_c$ the true rate prefactor should scale as a
(negative) power of $\epsilon$ for all $\epsilon\to 0$.  As in
\cite{MS93,MS96}, this can be treated quantitatively by studying the
'splayout' of the `tube' within which fluctuations are largely confined, as
the saddle is approached.  `Splayout' here simply means that the
fluctuational tube width, which for $\ell\ne\ell_c$ is $O(\epsilon^{1/2})$,
becomes $O(\epsilon^\alpha)$, with $\alpha<1/2$, as the saddle is
approached when $\ell=\ell_c$.  In the model studied in \cite{MS93,MS96},
the Lagrangian manifold comprising optimal fluctuational trajectories has a
more complicated behavior than in the model under study here.  As a result,
in \cite{MS93,MS96} fluctuations near the saddle occur on all lengthscales,
while in the present case fluctuations near the saddle occur on a definite
lengthscale, but one larger than $O(\epsilon^{1/2})$, leading to
non-Arrhenius behavior for all $\epsilon\to 0$.  Details will be presented
in \cite{MSinprep}.

\begin{figure}[t]
\centerline{\epsfig{file=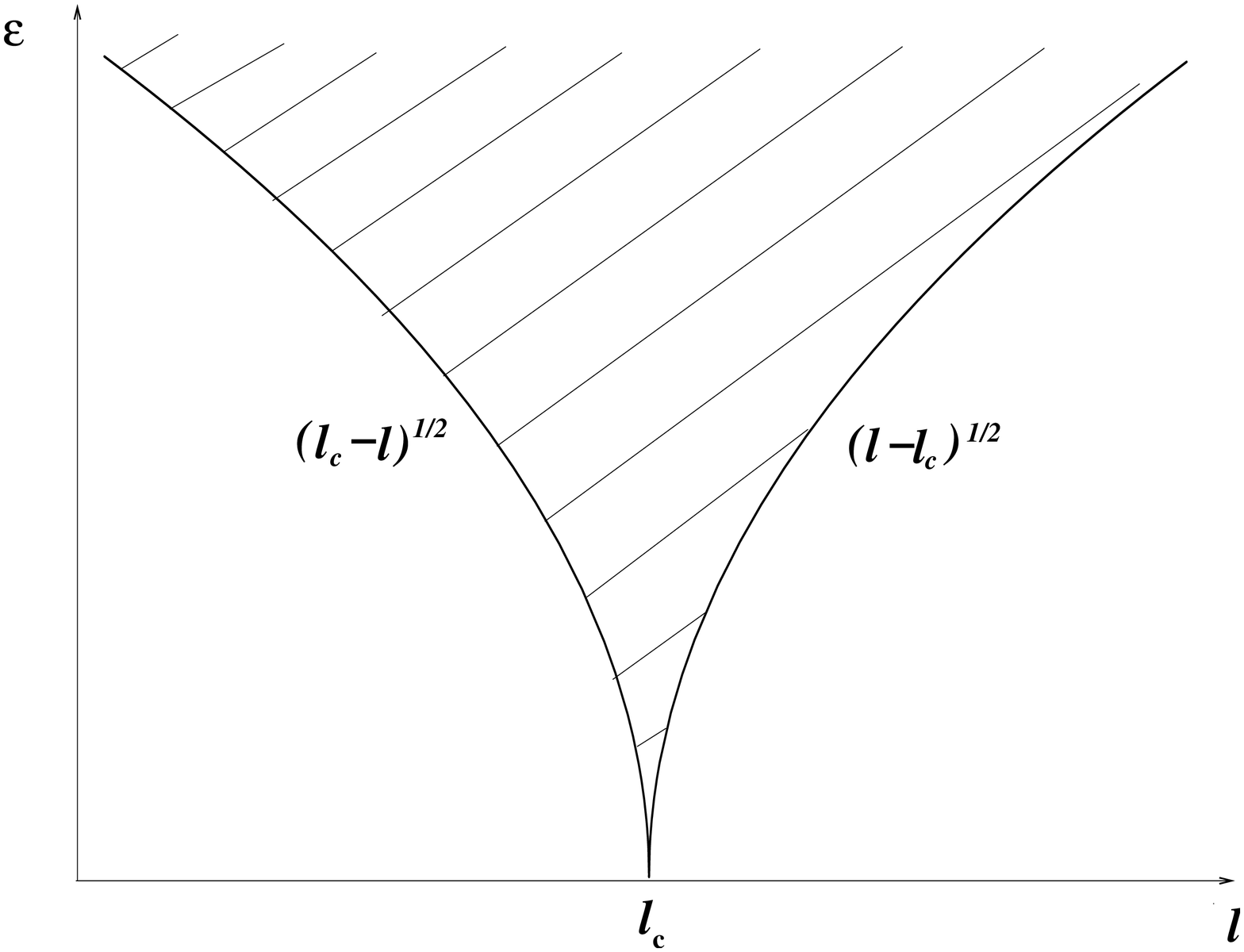,width=4.0in,height=3.0in}}
\renewcommand{\baselinestretch}{1.0} 
\small
\caption{A sketch of the scaling of the regions where the Arrhenius
prefactor formulae given by Eqs.~(\ref{eq:g0-}) and (\ref{eq:prefabove})
are valid, when $\ell$ is very close to $\ell_c$.  For fixed $\ell$,
$\epsilon$ must be small enough so that it is below the shaded region for
these formulae to apply.  Non-Arrhenius behavior is expected in the shaded
region.  Because this figure is intended to illustrate only the rate of
scaling of $\epsilon$ with $|l-l_c|$ so that the Kramers rate formula is
valid, the axes are unmarked (except for $\ell=\ell_c$, where the
$\epsilon$-range has shrunk to zero).}
\label{fig:scalingregion}
\end{figure}

Our main interest here is in the region close to $\ell_c$, where the rate
prefactor $\Gamma_0$ is growing anomalously large (but remains everywhere
finite for all $\ell$ strictly away from $\ell_c$).  The formulas given
here are valid for $\epsilon$ sufficiently small so that the contribution
from the quadratic fluctuations about the relevant extremal state of ${\cal
H}[\phi]$ dominates the action.  As long as all eigenvalues of
$\hat\Lambda[u_u]$ are nonzero (excluding the zero mode arising from
translational symmetry, which may be extracted), and the norms of the
corresponding eigenfunctions are bounded away from zero, the prefactor
formula applies, but in an $\epsilon$-region driven to zero as
$\ell\to\ell_c$ by the rate of vanishing of the eigenvalue(s) of smallest
magnitude.  Therefore, as $\ell\to\ell_c$ from either side, the Kramers
rate formula applies when $\epsilon$ scales to zero at least as fast as
$|\ell-\ell_c|^{1/2}$ (of course, the constraints already mentioned in
Sec.~\ref{sec:intro} must continue to hold as well).  More precisely, the
criterion considered here (which is necessary but not {\it a~priori\/}
sufficient) for Arrhenius behavior to hold on either side of $\ell_c$ is
that the noise strength $\epsilon$ be small compared to
$\lambda_m\langle\eta_m|\eta_m\rangle$, where $\lambda_m$ is the eigenvalue
of smallest magnitude and $\eta_m$ its corresponding eigenfunction(s).
This criterion arises from the condition, used in the derivation of
Eq.~(\ref{eq:prefactor}), that the noise strength be small compared to the
size of quadratic fluctuations about the extremal action.  For $\ell$
slightly below $\ell_c$, these quantities are given in
Sec.~\ref{subsec:subcrit}, and above $\ell_c$, by
Eqs.~(\ref{eq:eigenvalue3}) and (\ref{eq:eigenfunction3}).  The resulting
computation is straightforward and the resulting $\epsilon$-region is
sketched in Fig.~\ref{fig:scalingregion} (in the dimensionless units used
here, the coefficients of the scaling terms are of $O(1)$).

\renewcommand{\baselinestretch}{1.25} 
\normalsize 

The result is that, for fixed $\ell$ close to $\ell_c$, there should be a
{\it crossover\/} from non-Arrhenius to Arrhenius behavior at sufficiently
weak noise strength (cf.~\cite{MS96}).  The figure represents a type of
`Ginzburg criterion' that describes, at a given $\ell$ near $\ell_c$, how
far down as $\epsilon\to 0$ the non-Arrhenius behavior persists.  It should
be emphasized that Fig.~\ref{fig:scalingregion} sets an {\it upper bound\/}
on the scaling of the region of $\epsilon$ vs.~$|\ell-\ell_c|$, below which
asymptotic Arrhenius behavior sets in.  It would be interesting to consider
also the behavior at very small but fixed {\it noise\/} as $\ell$ increases
thorugh $\ell_c$. Here one would observe a crossover from Arrhenius to
non-Arrhenius behavior and back again as $\ell$ passes through the critical
region.  An interesting problem for future consideration is to analyze this
phenomenon in greater quantitative detail.

To summarize: strictly away from $\ell_c$, the prefactor formulae
Eqs.~(\ref{eq:g0-}) and (\ref{eq:prefabove}) hold (corresponding to
Arrhenius behavior of the rate), but in an increasingly narrow range of
$\epsilon$ as $\ell_c$ is approached.  A crossover from non-Arrhenius to
Arrhenius behavior as $\epsilon\to 0$ should be observed, along a boundary
that scales as shown in Fig.~\ref{fig:scalingregion}.  Strictly at
$\ell_c$, the formulae do not hold: the prefactor is finite, but acquires a
power-law (in $\epsilon$) character.  That is, the rate behavior is
non-Arrhenius all the way down to $\epsilon\to 0$.  This (boundary
condition-independent) non-Arrhenius behavior at criticality should be
distinguished from the (noncritical) non-Arrhenius behavior strictly above
$\ell_c$ that appears only when translation-invariant boundary conditions,
such as periodic, are used.

\section{Asymmetric quartic potentials}
\label{sec:asym}

\begin{figure}[t]
\centerline{\epsfig{file=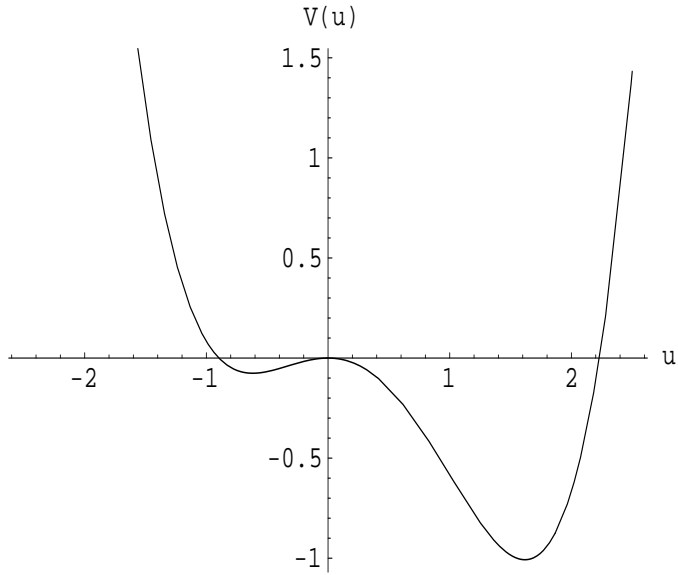,width=3.5in,height=3.0in}}
\renewcommand{\baselinestretch}{1.0} 
\small
\caption{The asymmetric quartic potential of Eq.~(\ref{eq:u4}).}
\label{fig:u4}
\end{figure}

In this section we will consider more general asymmetric potentials
of the form
\begin{equation}
\label{eq:u4}
V(u)=-(1/2)u^2-(1/3)u^3+(1/4)u^4
\end{equation}
as shown in Fig.~\ref{fig:u4}.

\renewcommand{\baselinestretch}{1.25} 
\normalsize 

We will consider only the small-$\ell$ regime, and show that a transition
at finite $\ell_c$ exists with a divergence of the prefactor as
$\ell\to\ell_c$.

Because the prefactor depends on the curvature of the potential near its
minimum, different prefactors (and of course activation energies)
correspond to the two minima $u_-=1/2-\sqrt{5}/2\approx -.62$ and
$u_+=1/2+\sqrt{5}/2\approx 1.62$.  The formulae shown here correspond to
$u_-$; corresponding formulae for escape from the potential minimum from
$u_+$ are obtained by replacing the constant $a_-=|1+2u_- -3u_-^2|\approx
1.38$ with $a_+=|1+2u_+ -3u_+^2|\approx 3.62$.  The critical length
$\ell_c$ is the same in the two cases.

\begin{figure}[t]
\vskip 0.5in
\centerline{\epsfig{file=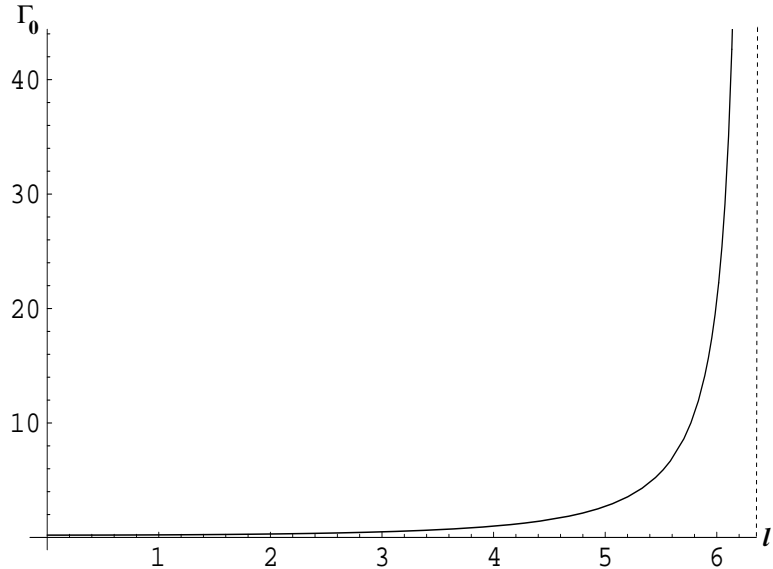,width=4.0in,height=3.0in}}
\renewcommand{\baselinestretch}{1.0} 
\small
\caption{The prefactor $\Gamma_0$ of Eq.~(\ref{eq:asympref}) for $\ell<\ell_c$.}
\label{fig:asympref}
\end{figure}

An analysis similar to that of Sec.~\ref{subsec:subcrit} yields
\begin{equation}
\label{eq:asympref}
\Gamma_0={(a_-)^{1/4}\over 2\pi}{\sinh(a_-\ell/2)\over\sin(\ell/2)}
\end{equation}
so $\ell_c=2\pi$ and, as before, the prefactor again diverges as
$(1-\ell/\ell_c)^{-1}$ as $\ell\to\ell_c^-$, as shown in
Fig.~\ref{fig:asympref}.

\renewcommand{\baselinestretch}{1.25} 
\normalsize 

\section{Conclusion}
\label{sec:conclusion}

We have found an explicit solution of the Kramers escape rate in an
asymmetric $\phi^3$ field theory of the Ginzburg-Landau form.  This result,
and the brief discussion in Sec.~\ref{sec:asym} of more general asymmetric
potentials, suggests that the critical behavior found in \cite{MS01} might
hold for a more general class of models than those with a high degree of
symmetry.  How widespread the transition phenomenon is remains uncertain,
but it appears to hold at least for arbitrary smooth potentials with terms
up to and including $\phi^4$.  It would be interesting to find models with
other types of behaviors.  One interesting possibility, discussed in
\cite{MS01}, is a class of models that display a {\it first\/}-order
transition: for example, a discontinuity in the derivative of the
activation barrier height with respect to the interval length, at a
critical length.  A possible candidate for such a model is the sixth-degree
Ginzburg-Landau potential of Kuznetsov and Tinyakov \cite{Kuznetsov97}, but
a detailed analysis of its transition behavior remains to be done.

\medskip

{\it Acknowledgment.}  I would like to thank Robert~Maier for useful
comments on the manuscript, and for many valuable discussions.

\renewcommand{\baselinestretch}{1.0} 

\end{document}